\documentclass[5p,twocolumn,times]{elsarticle}

\usepackage{lipsum}
\usepackage{lineno,hyperref}
\modulolinenumbers[5]
\usepackage[pdftex]{color}
\usepackage[font=footnotesize,labelfont=bf]{caption}
\usepackage[font=footnotesize,labelfont=bf]{subcaption}
\journal{Journal of \LaTeX\ Templates}

\usepackage{amssymb}

\biboptions{compress}

\usepackage[figuresright]{rotating}

\begin{document}
\begin{frontmatter}

\title{Ambush strategy impacts species predominance and coexistence in rock-paper-scissors models}


\address[1]{Research Centre for Data Intelligence, Zuyd University of Applied Sciences, Paul Henri Spaaklaan 3D, 6229 EN, Maastricht, The Netherlands}

\address[2]{Polytechnic School, Federal University of Bahia, Prof. Aristides Novis Street 2, 40210-910, Salvador, Bahia, Brazil}

\address[3]{Department of Computer Engineering and Automation, Federal University of Rio Grande do Norte, Av. Senador Salgado Filho 300, Natal, 59078-970, Brazil}

\author[1]{ J. Menezes} 
\author[2,3]{R. Barbalho} 

\begin{abstract}
We investigate the adaptive Ambush strategy in cyclic models following the rules of the spatial rock–paper–scissors game. In our model, individuals of one species possess cognitive abilities to perceive environmental cues and assess the local density of the species they dominate in the spatial competition for natural resources. Based on this assessment, they either initiate a direct attack or, if the local concentration of target individuals does not justify the risk, reposition strategically to prepare an ambush. To quantify the evolutionary consequences of these behavioural strategies, we perform stochastic simulations, analysing emergent spatial patterns and the dependence of species densities on the threshold used by individuals to decide between immediate attack or anticipation. Our findings reveal that, despite being designed to enhance efficiency, cognitive strategies can reduce the abundance of the species due to the constraints of cyclic dominance. We identify an optimal decision threshold: attacking only when the local density of target individuals exceeds $15\%$ provides the best balance between selection risk and long-term persistence. Furthermore, the Ambush strategy benefits low-mobility organisms, increasing coexistence probabilities by up to $53\%$. These results deepen the understanding of adaptive decision-making in spatial ecology, linking cognitive complexity to ecosystem resilience and extinction risk.
\end{abstract}

\end{frontmatter}

\section{Introduction}
\label{sec1}
The importance of spatial structure for maintaining biodiversity has been clearly demonstrated in microbial experiments with Escherichia coli \cite{bacteria}. In such systems, three competing bacterial strains survive through cyclic dominance, reflecting the dynamics of the rock–paper–scissors game: scissors defeat paper, paper defeats rock, and rock defeats scissors \cite{mobilia2,Avelino-PRE-86-036112,uneven,sara}. Coexistence is preserved only when competition takes place locally, producing spatial domains that protect diversity \cite{Coli}. Comparable mechanisms have been reported in other ecosystems, including Californian coral reef invertebrates and lizards in the inner Coast Range of California \cite{coral,lizards}. In these communities, restricted dispersal fosters localised interactions that stabilise diversity, whereas high levels of mobility homogenise the population and ultimately undermine coexistence \cite{Allelopathy}.

A substantial body of evidence indicates that organisms often adapt their behaviour in response to environmental cues \cite{ecology,Causes}. These behavioural adjustments are central to survival, persistence, and ecological stability \cite{MovementProfitable,Nature-bio}. For example, animals regularly alter their movement patterns to track resources, evade predators, or locate habitats suitable for reproduction. The capacity to sense and respond to local signals enables individuals to adjust their locomotion dynamically. Insights into these adaptive mobility strategies have also inspired advances in engineering, where robotic systems are designed to emulate animal movement and decision-making \cite{foraging,butterfly,BUCHHOLZ2007401,adaptive1,adaptive2,Dispersal,BENHAMOU1989375,coping}.

Among the behavioural strategies observed across taxa, ambush predation is rampant. This mode of foraging aims to maximise encounter success by striking at prey or competitors from concealment \cite{ambush0,ambush00}. Marine copepods provide a striking illustration: many species employ an ambush feeding mode in which sudden, high-speed jumps result in rapid attacks that leave prey little chance to escape \cite{ambush2}.
Ambush tactics are not uniform but depend on multiple ecological and physiological factors. It has been shown that predators may adjust their strikes based on prey characteristics such as body size, movement, and temperature \cite{ambush1}. For instance, vipers tend to attack larger prey, and their strike success is strongly influenced by both movement and thermal conditions, regardless of whether the prey is alive or dead. 

In the context of the rock-paper-scissors model, behavioural movement strategies play a vital role in species persistence and biodiversity maintenance \cite{tenorio1,tenorio2}. It has been shown that defensive strategic movement, which allows organisms to escape elimination by enemies (Safeguard strategy \cite{Moura}) or avoid infection during epidemic outbreaks (Social Distancing \cite{combination,adaptivej,adaptivejj,MENEZES2025105616}), is efficient at protecting individuals, benefiting the prevailing species in the spatial game.

The Ambush behavioural movement strategy, which allows individuals either to move directly toward organisms they aim to defeat in the spatial rock-paper-scissors  game, or to relocate in a different direction to wait and attack them upon arrival, was introduced in Ref. \cite{ambush1}. Using the May–Leonard implementation of the rock-paper-scissors  game, it was shown that when organisms of one species employ a locally adaptive Ambush strategy, their performance in the spatial game increases significantly. Moreover, the authors demonstrated that combining Attack and Anticipation movement strategies (first studied separately in \cite{Moura}) with a threshold based on the local density of organisms of the target species amplifies effects on species performance and spatial dynamics. Their simulations revealed that the perception radius $R$ critically influences success, with higher threshold triggers correlating with lower species densities. 

Here, we investigate the influence of the Ambush strategy on population dynamics and biodiversity when organisms possess the cognitive ability to implement it accurately. To this end, we perform large-scale simulations to quantify how species densities vary across different ambush thresholds. Finally, by considering organisms with varying mobility probabilities, we compute coexistence probabilities to determine whether the Ambush strategy promotes or jeopardises biodiversity.

Our analysis is conducted within the May–Leonard implementation of the rock-paper-scissors  model, where organisms engage in local interactions and population size is not conserved \cite{Menezes_2023,tanimoto2,Szolnoki-JRSI-11-0735,Anti1,anti2,MENEZES2022101606,PhysRevE.97.032415,Bazeia_2017,PhysRevE.99.052310}.

The outline of this paper is as follows. Section~\ref{sec2} introduces the model and methodology, including simulation details. Section~\ref{sec3} presents the impact of the proportion of individuals with ability to perform the Ambush strategy on spatial patterns. Section~\ref{sec4} explores the effects of the Ambush strategy on population dynamics and individual safety. Section~\ref{sec5} examines coexistence probabilities as a function of the ambush balance between Attack and Anticipation. Finally, Section~\ref{sec6} provides our discussion and conclusions.

\begin{figure}
\centering	
\includegraphics[width=40mm]{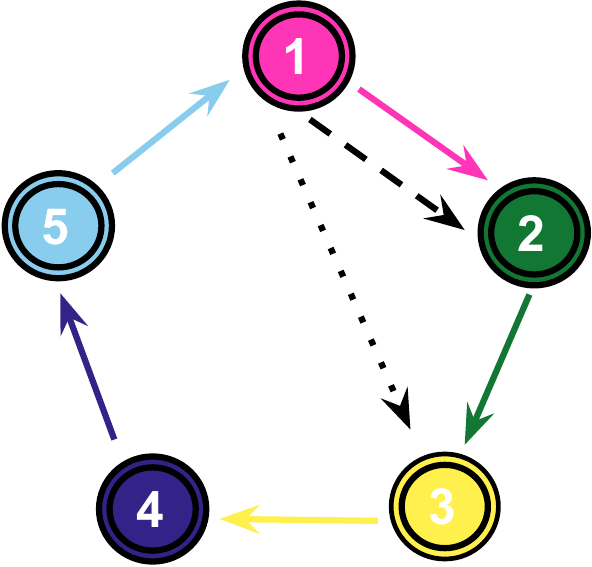}
\caption{Schematic representation of the five-species rock–paper–scissors model. Arrows indicate the dominance relations, where organisms of species $i$ outcompete those of species $i+1$. The two components of the Ambush strategy are highlighted: i) the dashed arrow (Attack) marks movements of species~$1$ toward regions with a higher density of species~$2$; ii) the dotted arrow (Anticipation) depicts displacements toward areas where species~$3$ are more abundant.}
\label{fig1}
\end{figure}
\begin{figure*}
	\centering
    \begin{subfigure}{.19\textwidth}
        \centering
        \includegraphics[width=34mm]{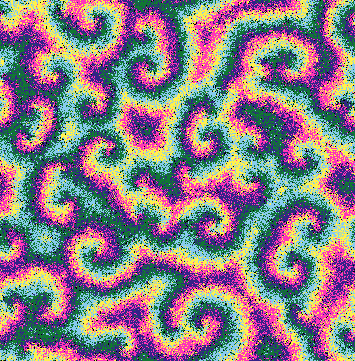}
        \caption{}\label{fig2a}
    \end{subfigure} %
   \begin{subfigure}{.19\textwidth}
        \centering
        \includegraphics[width=34mm]{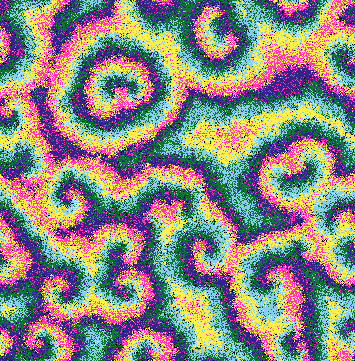}
        \caption{}\label{fig2b}
    \end{subfigure} 
            \begin{subfigure}{.19\textwidth}
        \centering
        \includegraphics[width=34mm]{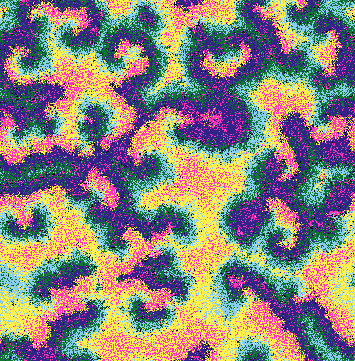}
        \caption{}\label{fig2c}
    \end{subfigure} 
           \begin{subfigure}{.19\textwidth}
        \centering
        \includegraphics[width=34mm]{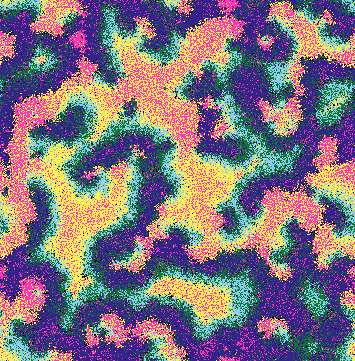}
        \caption{}\label{fig2d}
    \end{subfigure} 
   \begin{subfigure}{.19\textwidth}
        \centering
        \includegraphics[width=34mm]{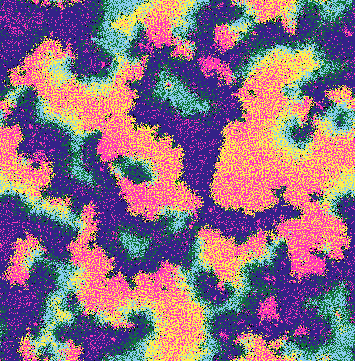}
        \caption{}\label{fig2e}
            \end{subfigure}
 \caption{Snapshots of final organims' spatial distribution captured from stochatic simulations of the five-species rock–paper–scissors model with Ambush Strategy. 
The lattice contains $500^2$ sites and evolves for $5000$ generations,
starting with the random initial configurations. Fig.~\ref{fig2a} shows the scenario where no organism can perform the behavioural movement, while Figs.~\ref{fig2b}, ~\ref{fig2c}, ~\ref{fig2d}, and ~\ref{fig2e}, the fraction of organisms of species $1$ performing the Ambush strategy is $25\%$, $50\%$, $75\%$ and $100\%$, respectively.
Colours follow the scheme of Fig.~\ref{fig1}, with empty sites represented by white dots.}
  \label{fig2}
\end{figure*}
\begin{figure}
    \begin{subfigure}{0.4\textwidth}
        \centering
\includegraphics[width=72mm]{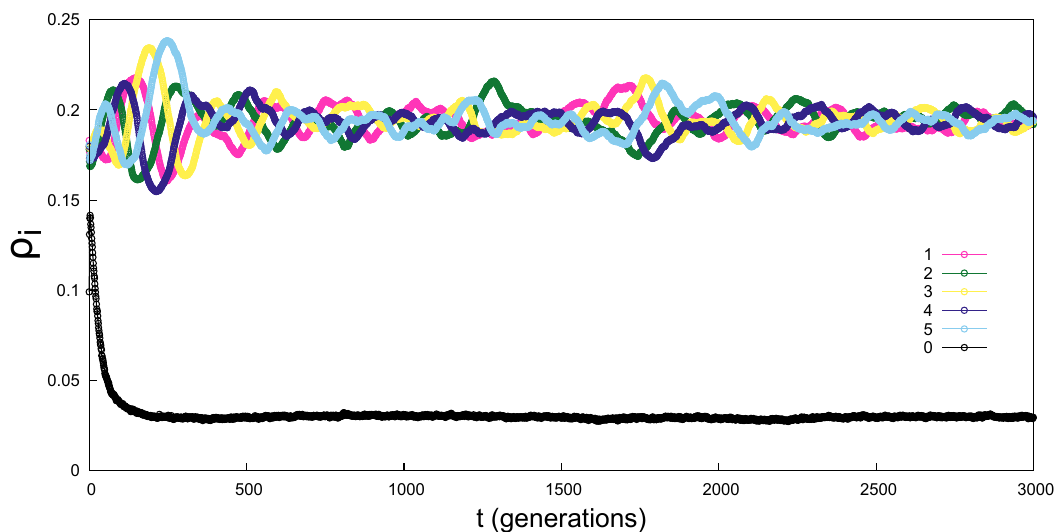}
        \caption{}\label{fig3a}
    \end{subfigure} %
   \begin{subfigure}{0.4\textwidth}
        \centering
\includegraphics[width=72mm]{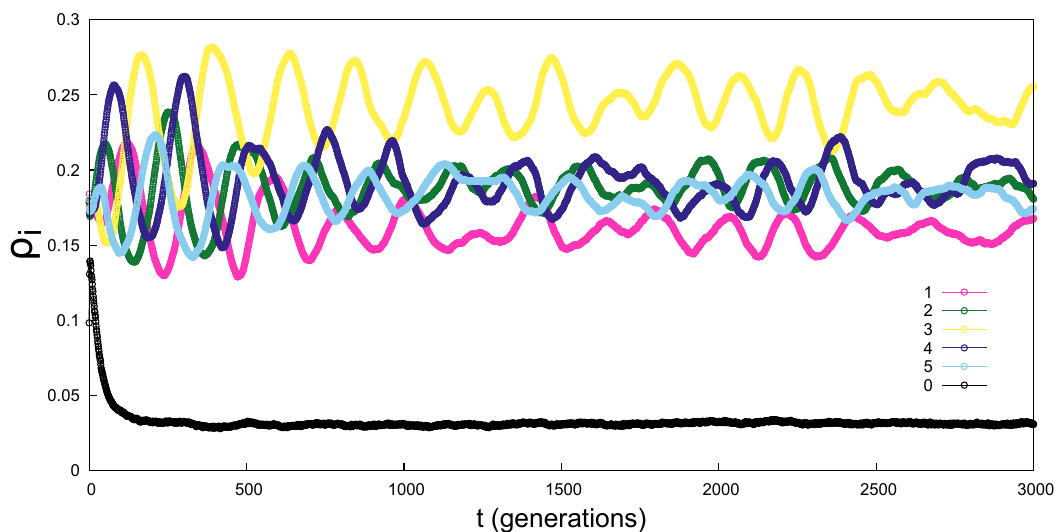}
        \caption{}\label{fig3b}
    \end{subfigure} 
            \begin{subfigure}{0.4\textwidth}
        \centering
\includegraphics[width=72mm]{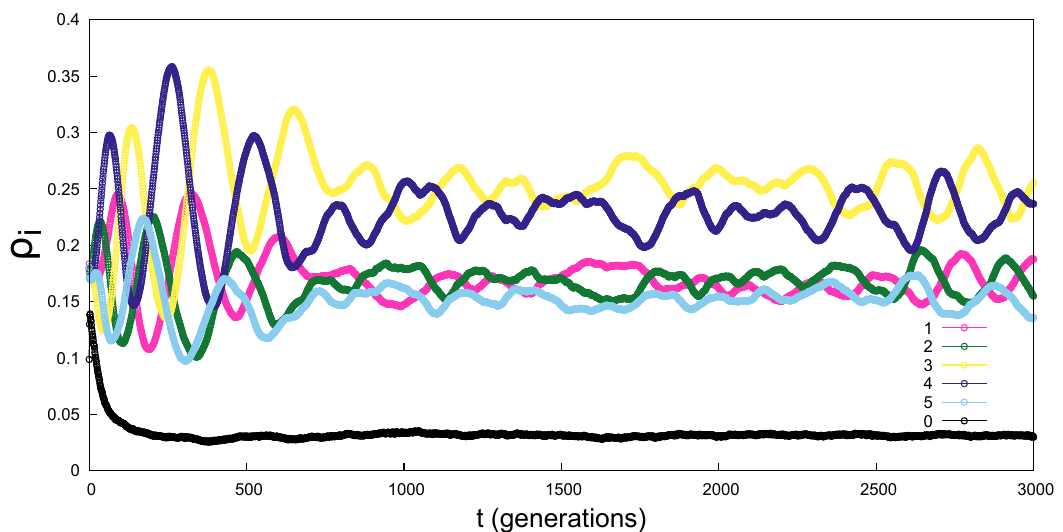}
        \caption{}\label{fig3c}
    \end{subfigure} 
           \begin{subfigure}{0.4\textwidth}
        \centering
  \includegraphics[width=72mm]{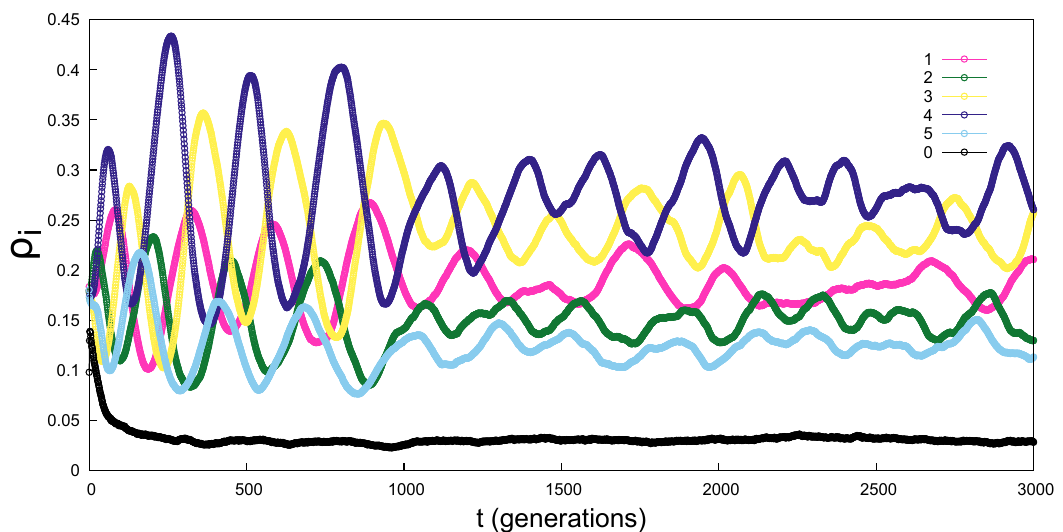}
        \caption{}\label{fig3d}
    \end{subfigure} 
   \begin{subfigure}{0.4\textwidth}
        \centering
\includegraphics[width=72mm]{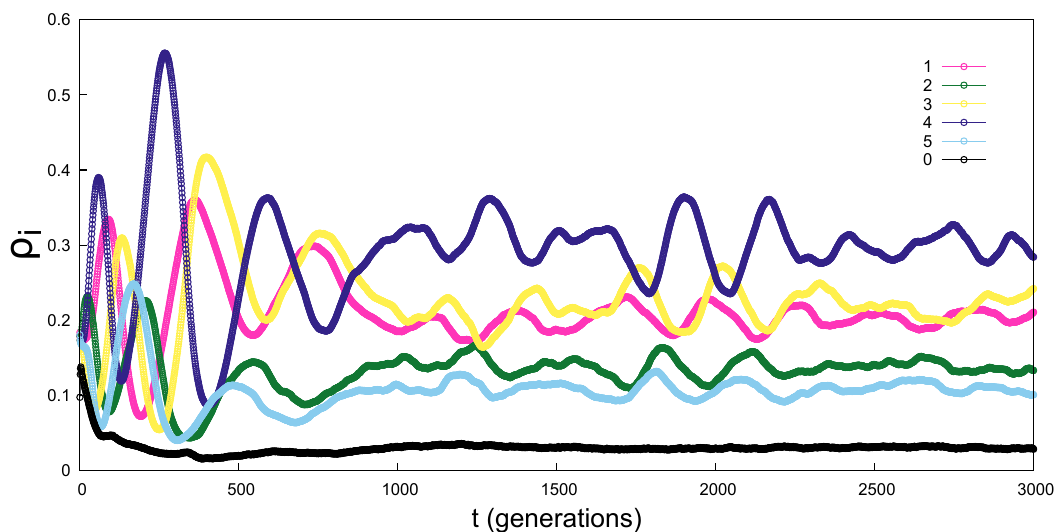}
        \caption{}\label{fig3e}
            \end{subfigure}
 \caption{Temporal dependence of the species densities for the simulations shown in Fig. \ref{fig2}. The black solid line depicts the density of empty spaces, while the pink, green, yellow, purple, and blue lines show the abundance of individuals of species $1$, $2$, $3$, $4$, and $5$, respectively. Figs. \ref{fig3a}, \ref{fig3b}, \ref{fig3c}, \ref{fig3d}, and \ref{fig3e} show how $\rho_i$ changes in the implementations whose final snapshot is shown in 
\ref{fig2a}, \ref{fig2b}, \ref{fig2c}, \ref{fig2d}, and \ref{fig2e}, respectively.
}
  \label{fig3}
\end{figure}
\section{Model and methods}
\label{sec2}
Our framework extends the spatial rock-paper-scissors model by incorporating a five-species cyclic competition structure, illustrated in Fig.~\ref{fig1}. Interactions follow the rule 
that organisms of species $i$ eliminate individuals of species $i+1$, with indices taken modulo five 
($i = i + 5\alpha$, $\alpha \in \mathbb{Z}$) \cite{ramylla}. This scheme generalises the classical 
three-species cycle, allowing us to explore how spatial heterogeneity and adaptive behaviours 
shape coexistence and population stability.  

We study the ambush-driven foraging strategy of species~$1$, 
designed to optimise competitive performance through real-time environmental assessment, as introduced in Ref.\cite{ambush}. Accordingly, Fig.~\ref{fig1} illustrates the two components of the Ambush strategy:  
(i) the dashed arrow (Attack) indicates movements of species~$1$ toward regions with higher densities of species~$2$;  
(ii) the dotted arrow (Anticipation) represents displacements toward areas where species~$3$ are more abundant.

The behaviour unfolds in three stages:  
\begin{enumerate}
    \item Local Environment Scan: Individuals of species~$1$ examine their surroundings 
    within a perception radius $R$ (in lattice units) to quantify the local density of individuals of species~$2$.  

    \item Attack Motion: If the local density of species~$2$ exceeds a threshold $\beta$ 
    ($0 \leq \beta \leq 1$), the organism of species $1$ executes the Attack strategy, moving toward the 
    sector of highest concentration of individuals of species $2$ \cite{Moura}.  

    \item Anticipatory Relocation: If prey density falls below $\beta$, the individual of species $1$ adopts 
    the \textbf{Anticipation strategy} \cite{Moura}, relocating toward regions where species~$3$ 
    (the species that organisms of species~$2$ will replace). 
    This positions individuals of species~$1$ to exploit future resource availability, representing 
    a rudimentary form of predictive foraging.  
\end{enumerate}

This adaptive tactic contrast to the random dispersal assumed for all 
other species, in line with standard spatial rock--paper--scissors modelling \cite{mobilia2}.  

We perform stochastic simulations on $N \times N$ square lattices with periodic boundaries; 
each site hosts at most one individual. Initial conditions assign equal population densities 
across all five species ($I_i = N^2/5$), following the non-conservative May--Leonard formulation 
\cite{leonard}. Interactions occur within von Neumann neighbourhoods (four nearest neighbours) 
and are governed by three stochastic processes:  

\begin{itemize}
    \item Selection: $i\ j \to i\ \otimes$ \quad (where $j = i+1$ and $\otimes$ denotes an empty site),  
    \item Reproduction: $i\ \otimes \to i\ i$,  
    \item Mobility: $i\ \odot \to \odot\ i$ \quad (with $\odot$ representing any species or vacancy).  
\end{itemize}

Rates $s$, $r$, and $m$ (selection, reproduction, and mobility) are set uniformly across species and normalised to $s = r = m = 1/3$, unless explicitly stated otherwise. Robustness checks confirm that the qualitative dynamics are preserved under parameter variation.

For species~$1$, directional movement is determined through two phases:  
\begin{enumerate}
    \item Strategy Selection:  
    \begin{itemize}
        \item A perception zone of radius $R$ is defined around the focal individual.  
        \item Local densities of species~$2$ are computed. If density $\geq \beta$, where $0 \leq \beta \leq 1$ Attack
        strategy is chosen; otherwise, Anticipation is triggered.  
    \end{itemize}

    \item Directional Bias:  
    \begin{itemize}
        \item The perception zone is partitioned into four sectors aligned with the lattice directions.  
        \item Target densities are tallied per sector: species~$2$ for Attack, species~$3$ for Anticipation.  
        \item Movement occurs toward the sector with maximal target density (with ties broken randomly).  
    \end{itemize}
\end{enumerate}

Simulations are performed on $500^2$ lattices and run for $5000$ generations, ensuring equilibration and statistical reliability of the outcomes. Following Ref.~\cite{ambush}, we assume $R=5$ throughout this study, unless explicitly stated otherwise.

\section{Spatial Patterns}
\label{sec3}
To investigate how the proportion of organisms that are physiologically and cognitively apt to
 performing the Ambush strategy affects emergent spatial organisation in cyclic competition,
 we run a series of single realisations to observe 
 the changes in the organisms' spatial distribution.
 
Figures~\ref{fig2a}--\ref{fig2e} presents final snapshots from a series of simulations starting from random initial conditions and running for $3000$ generations, with $0\%$,
 $25\%$, $50\%$, $75\%$, and $100\%$ of individual of species $1$ executing ambush, respectively (the remaining individuals move randomly).
 
As it is well established in the literature (see \cite{Moura}, for example), cyclic rock–paper–scissors dynamics in generalised models produce planar waves that propagate across the grid. Spatial domains of pairs such as ${3,5}$ (dark blue, light blue) are progressively
 replaced by ${2,5}$ (green, light blue), then ${2,4}$ (green, yellow), followed by ${1,4}$ (pink, yellow), ${3,4}$ (dark blue, yellow), and ${1,3}$ (pink, dark blue), repeating periodically
 under cyclic dominance. The paired species are those that do not attack each other in the spatial game. 
 
 Figure \ref{fig2a} shows that, if no organism of species $1$ can scan or interpret the environment to employ the movement strategy, spiral waves emerge because all individuals of every species move randomly. However, once a small fraction of organisms begin using Ambush tactics to improve their performance, the symmetric formation of spatial patterns starts to erode, as shown in Fig. \ref{fig2b} \cite{ambush1}. This effect intensifies as more and more organisms are apt to perform the Ambush strategy, with the spiral waves nearly disappearing as the proportion of species $1$ individuals performing Ambush approaches totality, as shown in Figs.~\ref{fig2c}--\ref{fig2e}. These patterns
 differ sharply from random-movement models, highlighting how adaptive decision-making
 can destabilise equilibrium structures.

To assess the impact of this asymmetry on species abundance, we define the spatial density of species $i$, $\rho_i$ (with $i=1,\dots,5$), as the fraction of the grid occupied by individuals of that species at time $t$:  
\begin{equation}
\rho_i = \frac{I_i}{\mathcal{N}},
\end{equation}
where $I_i$ is the number of individuals of species $i$, and $\mathcal{N}$ is the total number of lattice sites. The proportion of empty sites is denoted by $\rho_0$.

The temporal dependence of species densities for simulations in Fig.\ref{fig2a}--\ref{fig2e}, are depicted in Figs. \ref{fig3a}--\ref{fig3e}. Pink, green, yellow, purple, and blue lines track the densities of species $1$, $2$, $3$, $4$, and $5$, respectively, while the black line represents empty spaces.
As expected, the average densities of all species remain balanced in the standard model, as shown in Fig. \ref{fig3a}. As more organisms of species $1$ move strategically, however, species densities become imbalanced, with the dominant species depending on the fraction of organisms of species $1$ adopting the Ambush strategy. For small proportion of individuals performing the Ambush strategy, as in Fig.~\ref{fig3b}, species $3$ remains the most abundant throughout the simulation. The dominance of species $3$ weakens as more organisms of species $1$ move strategically, with species $4$ gaining prominence, as indicated in Figs. \ref{fig3c} and \ref{fig3d}. Finally, for more than $50\%$ individuals executing Ambush strategy, species $4$ becomes dominant, with its average density increasing as more organisms of species $1$ employ the Ambush strategy, as shown in Figs. \ref{fig3d} and \ref{fig3e}.




\section{Selection Risk and Species Densities}
\label{sec4}

The Ambush strategy efficiently enhances individual performance in cyclic spatial games, as shown in Ref.~\cite{ambush}. However, because our framework is cyclic, improved efficiency in eliminating an opponent reverberates through the entire ecosystem, ultimately affecting the safety of the species adopting the tactic, as highlighted in Ref.~\cite{Moura}. Specifically, when individuals of species~$1$ evolve to perform the Ambush strategy, their success rate in killing species~$2$ increases, leading to a decline in the population of species~$2$. As a result, species~$3$ prospers, since fewer predators (species~$2$) remain to suppress it. This shift propagates further: species~$4$ and $5$ are also impacted, with the cascading consequences feeding back to species~$1$.

\begin{figure}[t]
   \centering
  \includegraphics[width=85mm]{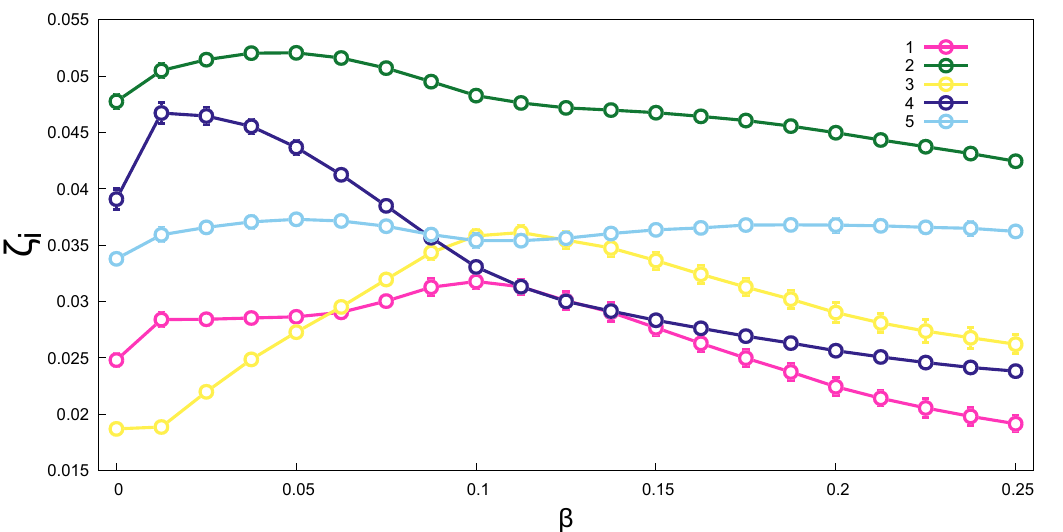}
  \caption{Changes in the species selection risk due to the Ambush movement strategy in terms of the Attack trigger. The outcomes were averaged
from sets of $100$ simulations, starting from different initial conditions in lattices with $500^2$ grid sites, running until $5000$ generations. 
The error bars show the standard deviation. The colours follow the scheme in Fig.\ref{fig1}. The cognitive factor is $\beta=1$, while the interaction parameters are set to $s=r=m=1/3$.}
 \label{fig4}
\end{figure}

\begin{figure}[t]
   \centering
  \includegraphics[width=85mm]{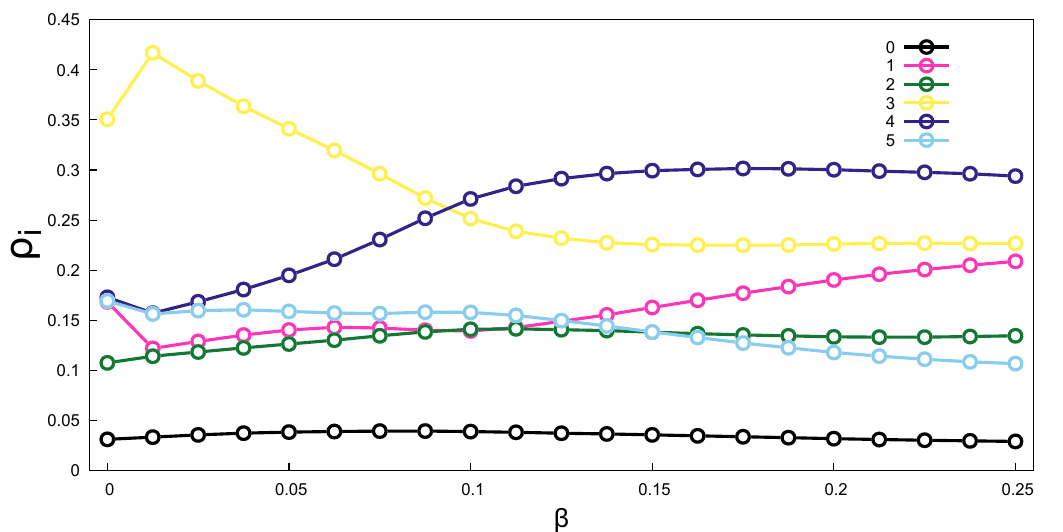}
  \caption{Changes in the species densities due to the Ambush movement strategy in terms of the Attack trigger. The outcomes were averaged
from sets of $100$ simulations, starting from different initial conditions in lattices with $500^2$ grid sites, running until $5000$ generations. 
The error bars show the standard deviation. The colours follow the scheme in Fig.\ref{fig1}. The cognitive factor is $\beta=1$, while the interaction parameters are set to $s=r=m=1/3$.}
 \label{fig5}
\end{figure}

To analyse how the adaptive Ambush tactic influences population dynamics, we calculate the mean species densities $\rho_i$ and  
examine the effect of adaptive movement on the selection risk of species $i$, denoted by $\zeta_i$. The algorithm proceeds as follows:  
\begin{enumerate}
\item count the total number of individuals of species $i$ at the beginning of each generation;  
\item compute the number of individuals of species $i$ eliminated during the generation;  
\item calculate $\zeta_i$ as the ratio between the eliminated individuals and the initial population.  
\end{enumerate}
Figures~\ref{fig4} and \ref{fig5} show $\zeta_i(\%)$ and $\rho_i(\%)$ as functions of the attack trigger $\beta$. Circles represent mean values, while error bars indicate standard deviations.

Figure~\ref{fig4} show that the selection risk of species~$2$ is the highest among all species, regardless of the attack trigger $\beta$. Accordingly, its population remains the smallest, as shown by the green curve in Fig.~\ref{fig5}. Interestingly, as $\beta$ increases, $\rho_2$ also rises, and for $\beta \geq 0.15$, the density of species~$2$ surpasses that of species~$5$. This finding confirms the results of Fig.~\ref{fig3}: when few organisms prepare ambushes, species~$3$ dominates, whereas at higher ambush adoption rates, species~$4$ becomes the prevailing species.

The results also reveal that this transition in the attack trigger critically affects species~$1$, the Ambush strategists. For $\beta > 0.1$, their selection risk decreases, but only for $\beta > 0.15$ does the benefit translate into population growth. In other words, if individuals of species~$1$ attack only when the local density of species~$2$ exceeds $15\%$, their population expands more than if they attack indiscriminately 
without preparing ambushes.

\section{Coexistence Probability}
\label{sec5}
\begin{figure}[t]
   \centering
  \includegraphics[width=85mm]{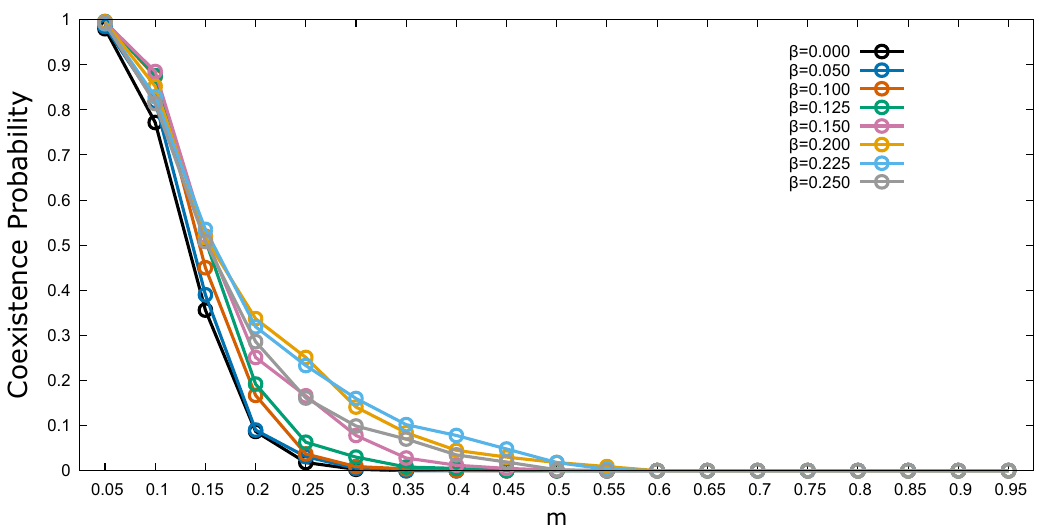}
  \caption{Coexistence probability as a function of the mobility $m$ for the generalised
rock–paper–scissors with five species. The black, dark blue, orange, light green, dark pink, orange, cyan, and grey lines in Fig.\ref{fig6}, respectively for $\beta=0.0$, $\beta=0.05$, $\beta=0.1$, $\beta=0.125$, $\beta=0.15$, $\beta=0.2$, $\beta=0.225$, and $\beta=0.25$. 
The interaction parameters are given by $s = r = (1-m)/2$.
The outcomes show the fraction of simulations ending in coexistence in collections of $1000$ implementations using lattices with $100^2$ grid sites, running until $10000$ generations. }
 \label{fig6}
\end{figure}

To investigate the effects of the Ambush strategy on biodiversity, we calculate the coexistence probability. For this purpose, we performed $1000$ simulations on $100^2$ lattices, each running for $10000$ generations, across a wide range of mobility probabilities $m$.  
Each simulation started from random initial conditions. Coexistence is defined as the condition in which at least one individual of every species is present at the end of the simulation, while extinction is recorded whenever at least one species is absent.
Therefore, the coexistence probability corresponds to the fraction of realisations resulting in coexistence. 
To quantify coexistence in terms of mobility, we repeated the simulations for $0.05 \leq m \leq 0.95$ in steps of $\Delta m = 0.05$, setting the selection and reproduction probabilities to $s = r = (1-m)/2$. The results are presented in Fig.~\ref{fig6}.  

In general, the probability that all species survive until the end of the simulation decreases with increasing $m$, regardless of the behavioural tactic adopted \cite{mobilia2,Moura}. The baseline case in which organisms always attack without restriction ($\beta=0$) is shown by the black line. The results indicate that indiscriminate attacking is the least favourable strategy: biodiversity is lost whenever organisms move with a probability greater than $30\%$. By contrast, implementing the Ambush strategy promotes biodiversity by increasing the likelihood that all species persist. This effect is depicted by the dark blue, orange, light green, dark pink, orange, cyan, and grey lines in Fig.~\ref{fig6}, corresponding respectively to $\beta=0.05$, $\beta=0.1$, $\beta=0.125$, $\beta=0.15$, $\beta=0.2$, $\beta=0.225$, and $\beta=0.25$.

Our findings show that, overall, biodiversity is enhanced for every $\beta>0$. Interestingly, the results also reveal that small values of $\beta$, where fewer individuals anticipate and prepare ambushes, pose a greater threat to coexistence, while larger values of $\beta$ support biodiversity even when dispersal is nearly twice the critical threshold observed for pure attack ($\beta=0$). Moreover, the simulations demonstrate the existence of an optimal $\beta$ that maximises biodiversity promotion, depending on the dispersal rate of organisms. For instance, within the interval $0.15 < m < 0.3$, the choice of $\beta=0.2$ produces the most significant biodiversity gain, with coexistence probability increasing by approximately $23\%$ for $m=0.25$. Moreover, the greatest increase is found in the same mobility range, where the optimal attack trigger is $\beta=0.225$. In this case, the probability of maintaining biodiversity when organisms employ the Ambush strategy reaches $53\%$ for $m=0.3$.  


\section{Discussion and Conclusions}
\label{sec6}
Our study addresses a cyclic five-species system where organisms of species $i$ may have the ability to intelligently move toward directions that maximise their chances of success in the spatial rock–paper–scissors scenario. This behavioural strategy allows organisms of species~$1$ to detect regions with a high concentration of individuals of species~$i+1$ and decide whether the local density of targets is sufficient to ensure a successful invasion. In the affirmative case, they move toward that direction rather than wandering randomly. Otherwise, individuals position themselves strategically to intercept opponents likely to dominate the region.

We performed stochastic simulations to quantify the impact of the Ambush strategy adopted by species~$i$ on spatial pattern formation, assuming that only a fraction of organisms are able to perform the strategy. Our results show that the symmetric spiral waves generated by cyclic dynamics in classical rock–paper–scissors models gradually erode under Ambush behaviour. We also found that, as the game becomes imbalanced through the evolution of species~$i$, species~$i+2$ dominates when only a minority of organisms move strategically. In contrast, when more than half of the organisms employ the Ambush movement, species~$i+3$ becomes predominant.

Our findings further reveal that implementing Ambush strategies, which enhance the likelihood of individual success compared to the pure attack tactic \cite{Moura}, also modifies the selection risk of organisms across all species. Due to the cyclic symmetry of the rock–paper–scissors game, the optimum attack trigger is achieved for $\beta > 0.15$. Thus, if individuals of species~$i$ attack only when the local density of species~$i+1$ exceeds $15\%$, they reach the best balance between improving performance, reducing selection risk, and increasing overall population density. This result sheds light on ecological studies of animal behaviour and may help biologists to understand how decision-making processes optimise competition for natural resources while balancing individual safety and species persistence.

By computing the coexistence probability, we also demonstrate that Ambush behaviour represents an evolutionary strategy in terms of biodiversity preservation, in contrast with the innate behaviour of attacking whenever target organisms are encountered. Moreover, our results show that, in addition to raising coexistence probability, the choice of an appropriate threshold for deciding whether the local density of species~$i+1$ is sufficient to trigger an attack is crucial, especially for individuals dispersing with different mobility probabilities. According to our simulations, this choice can increase biodiversity preservation by up to $53\%$ for low-mobility individuals.

Beyond the Ambush dynamics examined here, our modelling framework offers a natural avenue to incorporate alternative behavioural strategies. One example is the safeguard strategy \cite{Moura}, in which individuals weigh the trade-off between aggression and self-preservation. Instead of always attacking, an organism may refrain from confrontation when environmental cues indicate that the risks of predation exceed the expected gains. Such context-dependent behaviour highlights how decision-making based on threat assessment can extend survival, alter spatial organisation, and ultimately reshape the evolutionary outcomes of cyclic rock–paper–scissors interactions.

\section*{Acknowledgments}
We thank CNPq/Brazil, ECT/UFRN, FAPERN/RN, IBED/UvA, ADSAI/Zuyd, and Brightlands Smart Services Campus for financial and technical support.

\bibliographystyle{elsarticle-num}
\bibliography{ref}

\end{document}